\documentclass[10pt,twocolumn,twoside]{IEEEtran}
\usepackage{amsmath}
\usepackage{amsfonts}
\usepackage{amssymb}
\usepackage{algorithm}
\usepackage{algorithmic}
\usepackage{graphicx}
\usepackage{verbatim}

\newtheorem{theorem}{Theorem}
\newtheorem{lemma}{Lemma}
\newtheorem{remark}{Remark}

\newtheorem{example}{Example}

\newcommand{\beq}{\begin{equation}}
\newcommand{\eeq}{\end{equation}}
\newcommand{\beqnn}{\begin{equation*}}
\newcommand{\eeqnn}{\end{equation*}}
\newcommand{\beqy}{\begin{eqnarray}}
\newcommand{\eeqy}{\end{eqnarray}}
\newcommand{\beqynn}{\begin{eqnarray*}}
\newcommand{\eeqynn}{\end{eqnarray*}}
\newcommand{\bit}{\begin{itemize}}
\newcommand{\eit}{\end{itemize}}
\newcommand{\ben}{\begin{enumerate}}
\newcommand{\een}{\end{enumerate}}
\newcommand{\bex}{\begin{example}}
\newcommand{\eex}{\end{example}}

\newcommand{\balg}[1]{\begin{algorithm} \caption{#1}}
\newcommand{\ealg}{\end{algorithm}}

\newcommand{\balgc}{\begin{algorithmic}[1]}
\newcommand{\ealgc}{\end{algorithmic}}

\newcommand{\bary}{\begin{array}}
\newcommand{\eary}{\end{array}}
\newcommand{\bmx}{\begin{bmatrix}}
\newcommand{\emx}{\end{bmatrix}}
\newcommand{\bsmx}{\left[\begin{smallmatrix}}
\newcommand{\esmx}{\end{smallmatrix}\right]}
\newcommand{\bmxc}[1]{\left[\begin{array}{@{}#1@{}}}
\newcommand{\emxc}{\end{array}\right]}
\newcommand{\bcn}{\begin{center}}
\newcommand{\ecn}{\end{center}}

\newcommand{\Rbb}{{\mathbb{R}}}

\newcommand{\Rnbn}{\Rbb^{n \times n}}

\newcommand{\Rmbm}{\Rbb^{m \times m}}

\newcommand{\A}{\boldsymbol{A}}

\newcommand{\I}{\boldsymbol{I}}

\renewcommand{\L}{\boldsymbol{L}}

\newcommand{\Q}{\boldsymbol{Q}}
\newcommand{\R}{\boldsymbol{R}}

\newcommand{\Z}{\boldsymbol{Z}}

\renewcommand{\v}{\boldsymbol{v}}

\newcommand{\x}{{\boldsymbol{x}}}
\newcommand{\y}{{\boldsymbol{y}}}
\newcommand{\z}{\boldsymbol{z}}
\newcommand{\0}{{\boldsymbol{0}}}

\newcommand{\br}{{\bar{r}}}

\newcommand{\bby}{{\bar{\y}}}

\newcommand{\bbQ}{{\bar{\Q}}}
\newcommand{\bbR}{{\bar{\R}}}

\newcommand{\tby}{{\tilde{\y}}}

\newcommand{\hbx}{{\hat{\x}}}

\newcommand{\hbz}{{\hat{\z}}}

\newcommand{\bxi}{\boldsymbol{\xi}}

\usepackage{slashbox}

\begin{document}

\title{Effects of Some Lattice Reductions on the Success Probability of the Zero-Forcing Decoder}

\author{Jinming Wen, Chao~Tong, and Shi Bai 
\thanks{Jinming Wen is with CNRS, Laboratoire LIP (U. Lyon, CNRS, ENSL, INRIA, UCBL), Lyon, 69007, France (e-mail: jinming.wen@ens-lyon.fr).}
\thanks{Chao~Tong is with School of Computer Science and Engineering, Beihang University,  Beijing, 100191,  China. (e-mail: tongchao@buaa.edu.cn).}
\thanks{Shi Bai is with ENS de Lyon, Laboratoire LIP (U. Lyon, CNRS, ENSL, INRIA, UCBL), Lyon, 69007, France (e-mail: shi.bai@ens-lyon.fr).}
\thanks{This research  was supported by  ``Programme Avenir
Lyon Saint-Etienne de l'Universit\'e de Lyon" in the framework of the programme
``Inverstissements d'Avenir" (ANR-11-IDEX-0007),  ANR through the HPAC project under
Grant ANR~11~BS02~013, NSFC (No. 61472024), and ERC Starting Grant ERC-2013-StG-335086-LATTAC.}
}


%


\maketitle

\begin{abstract}
Zero-forcing (ZF) decoder is a commonly used approximation solution of the integer least squares problem which arises in communications and many other applications.
Numerically simulations have shown that the LLL reduction can usually improve the success probability $P_{ZF}$ of the ZF decoder.
In this paper, we first rigorously show that both SQRD and V-BLAST, two commonly used lattice reductions, have no effect on $P_{ZF}$.
Then, we show that LLL reduction can improve $P_{ZF}$ when $n=2$, we also analyze how the parameter $\delta$ in the LLL reduction affects the enhancement of $P_{ZF}$.
Finally, an example is given which shows that the LLL reduction decrease $P_{ZF}$ when $n\geq3$.
\end{abstract}


\section{Introduction}
In many applications including wireless communications (see e.g., \cite{DamGC03}), we need to detect
an unknown integer parameter vector $\hbx\in\mathbb{Z}^n$ from the following linear model
\begin{equation}
\label{e:model}
\y=\A\hbx+\v,
\end{equation}
where $\y\in \mathbb{R}^m$ is an observation vector,   $\A\in\mathbb{R}^{m\times n}$ is a deterministic model matrix with full column rank,
and $\v\in \mathbb{R}^m$ is a noise vector following the Gaussian distribution
$\mathcal{N}(\boldsymbol{0},\sigma^2 \I)$ with $\sigma$ being known.

A common method to estimate $\hbx$ in \eqref{e:model} is to solve the following integer least squares (ILS) problem:
\beq
\label{e:ILS}
\min_{\x\in\mathbb{Z}^n}\|\y-\A\x\|_2^2,
\eeq
whose solution is the maximum likelihood estimator of $\hbx$.

A typical approach to solving \eqref{e:ILS} is the discrete search approach,  referred to as sphere decoding in communications,
such as the Schnorr-Euchner algorithm \cite{SchE94} or its variants, see e.g. \cite{AgrEVZ02,DamGC03}.
To make the search faster,  a lattice reduction is usually performed to transform the given problem to an equivalent problem.
A widely used  reduction is the LLL reduction  proposed by Lenstra, Lenstra and Lov\'{a}sz in \cite{LenLL82}.
Another two  popular lattice reductions are SQRD \cite{WubBRKK01} and V-BLAST \cite{FosGVW99}.

The ILS problem \eqref{e:ILS} is NP-hard \cite{Boa81, Mic01}, so for some real-time applications, an approximate solution,
which can be produced quickly, is  computed instead. Two often used approximate solutions are the Zero-Forcing (ZF)
and the Successive Interference Cancelation (SIC) decoders,
which are respectively produced by Babai's rounding off and nearest plane algorithms \cite{Bab86}.
ZF decoder is one of the commonly used suboptimal estimators in communications, see, e.g., \cite{WubSJM11} and references therein,
and it has attracted much attention, see, e.g., \cite{TahMK07}.

In order to see how good an estimator is,
one needs to find the probability of the estimator being equal to the true integer parameter vector $\hat{x}$,
which is referred to as success probability \cite{HasB98}.

The success probability of the ZF decoder is very important because if it is close to 1,
we are confident to use it as a suboptimal solution of \eqref{e:ILS}, so we do not need to spend extra computational time to solve \eqref{e:ILS} to find the optimal estimator.
However, if it is low, then we need to use other more effective suboptimal solutions, or solve \eqref{e:ILS}.
SQRD \cite{WubBRKK01}, V-BLAST \cite{FosGVW99} and the LLL reduction \cite{LenLL82} are three most commonly used lattice reductions in sphere decoding.
Investigating their effects on the success probability $P_{ZF}$ of the ZF decoder is of vital importance
because if we know a lattice reduction can usually or always improve $P_{ZF}$, we can use it to improve the effectiveness of the ZF decoder,
otherwise, i.e., if a lattice reduction has no effect on $P_{ZF}$ or usually decreases $P_{ZF}$, then we do not need to spend time to do the reduction.
Numerical experiments have shown that after the LLL reduction, the success probabilities of the ZF and SIC decoders usually increase \cite{GanM08}.
How these lattice reductions affect the success probability of the SIC decoder has been investigated in \cite{ChaWX13} and \cite{WenC14}.
But how they affect the success probability of the ZF decoder is still unknown.
Since the formula for computing the success probability $P_{SIC}$ of the SIC decoder is totally different from that for the ZF decoder,
the method for studying the effects of lattice reductions on $P_{SIC}$  can not be applied to $P_{ZF}$.
In this paper, we aim to theoretically investigate the effects of SQRD, V-BLAST and LLL on the success probability of the ZF decoder
by using a new method.
Different from the SIC decoder, which is the permutation strategies that improve its success probability,
in this paper, we will show that the size reductions improve $P_{ZF}$.
If permutation strategies in the LLL reduction can bring more size reductions, certainly they may further increase $P_{ZF}$.

The rest of the paper is organized as follows.
In Section \ref{s:lattice}, we introduce lattice reductions to reduce the ILS problem \eqref{e:ILS}.
In Section \ref{s:effects}, we first introduce a formula to compute the success probability $P_{ZF}$ of the ZF decoder.
Then, we show that neither SQRD nor V-BLAST can change $P_{ZF}$, but LLL reduction can improve it when $n=2$.
After this, we analyze how the parameter $\delta$ in the condition of the LLL reduction affects the enhancement of $P_{ZF}$.
In the end, an example is given which shows that the LLL reduction decrease $P_{ZF}$ when $n\geq3$.
Finally,  we summarize this paper in Section \ref{s:sum}.

{\it Notation.}
Let $\mathbb{R}^n$ and $\mathbb{Z}^n$ be the spaces of the $n$-dimensional column real vectors and integer vectors, respectively.
Let $\mathbb{R}^{m\times n}$ and $\mathbb{Z}^{m\times n}$ be the spaces of the $m\times n$ real matrices and integer matrices, respectively.
Boldface lowercase letters denote column vectors and boldface uppercase letters denote matrices.
For a matrix $\A$, let $a_{ij}$ be the element at row  $i$ and column  $j$.
For a vector $\x$, let $x_{i}$ be its $i$-th element, and let $\lfloor \x\rceil$ to denote its nearest integer vector, i.e.,
each entry of $\x$ is rounded to its nearest integer (if there is a tie, the one with smaller magnitude is chosen).
The success probability of the ZF decoder is denoted by $P_{ZF}$.

\section{Lattice Reductions and transformation of the ILS Problem}\label{s:lattice}

Assume that $\A$ in the linear model \eqref{e:model} has the QR factorization
\beq
\label{e:qr}
\A=[\Q_1, \Q_2]\bmx\R \\ \0 \emx,
\eeq
where $[\underset{n}{\Q_1}, \underset{m-n}{\Q_2}]\in \Rmbm$ is orthonormal and $\R\in \Rnbn$ is upper triangular.
Without loss of generality, we assume $r_{ii}>0$ for $1\leq i\leq n$.

Define $\tby=\Q_1^T\y$.
From \eqref{e:model} and \eqref{e:qr},  we have
$$
\tby=\R\hbx+\Q_1^T \v.
$$
Since $\v\sim \mathcal{N}(\boldsymbol{0},\sigma^2 \I)$,
we obtain $\tby \sim \mathcal{N}(\R\hbx, \sigma^2 \I)$.
Moreover, the ILS problem \eqref{e:ILS} can be transformed to
\beq
\label{e:ILSR}
\min_{\x\in\mathbb{Z}^n}\|\tby-\R\x\|_2^2.
\eeq
One can then apply a sphere decoder such as the Schnorr-Euchner search algorithm \cite{SchE94} to solve \eqref{e:ILSR}.
For efficiency, before the search, one uses a lattice reduction to reduce $\R$.
After the QR factorization \eqref{e:qr} of $\A$, lattice reductions reduce $\R$ in \eqref{e:qr} to $\bbR$ by:
\beq
\label{e:QRZ}
\bbQ^T\R \Z = \bbR,
\eeq
where $\bbQ  \in \mathbb{R}^{n\times n}$ is orthogonal, $\bbR\in \mathbb{R}^{n\times n}$ is an upper triangular matrix
satisfying the properties of the corresponding reductions
and $\Z$ is a  unimodular matrix (i.e., $\Z  \in \mathbb{Z}^{n\times n}$ and $\det(\Z)=\pm1$).
In particular, if SQRD or V-BLAST is used to get \eqref{e:QRZ}, then $\Z$ is a permutation matrix.
After the lattice reduction \eqref{e:QRZ}, the ILS problem \eqref{e:ILSR} is then transformed to:
\beq
\label{e:reduced}
\min_{\z\in\mathbb{Z}^n}\|\bby-\bbR\z\|_2^2,
\eeq
where $\bby=\bbQ^T\tby$ and $\z=\Z^{-1}\x$.

The SQRD strategy determines the columns of $\bbR$ we seek from the first to the last by using the modified Gram-Schmidt method.
In the $k$-th step of the modified Gram-Schmidt method, the $k$-th column of the permuted $\bbR$ we seek is chosen from the remaining $n-k+1$ columns of $\R$  such that $|\br_{kk}|$  is smallest. For more details and efficient algorithms, see, e.g. \cite{WubBRKK01}.

In contrast to the SQRD strategy, the V-BLAST determines the columns of  $\bbR$ from the last to the first.
Suppose columns $k+1, k+2, \ldots, n$ of $\bbR$ have been determined, this strategy chooses a column from the remaining
columns of $\R$ as the $k$-th column such that $|\br_{kk}|$  is maximum over all $k$ choices.
For more details and efficient algorithms, see \cite{FosGVW99} and \cite{ChaP07}.

If the LLL reduction is used to get \eqref{e:QRZ}, then $\bbR$ satisfies the following two conditions:
\begin{align}
&|\br_{ik}|\leq\frac{1}{2} |\br_{ii}|, \quad i=1, 2, \ldots, k-1, \label{e:criteria1} \\
&\delta\, \br_{k-1,k-1}^2 \leq   \br_{k-1,k}^2+ \br_{kk}^2,\quad k=2, 3, \ldots, n, \label{e:criteria2}
\end{align}
where $\delta$ is a constant satisfying $1/4 < \delta \leq 1$.
The matrix $\A\Z$ is said to be  LLL reduced.
Equations \eqref{e:criteria1} and  \eqref{e:criteria2} are referred to as the size-reduced condition and the Lov\'{a}sz condition, respectively.

The original LLL reduction algorithm is described in Algorithm \ref{a:LLL}, where
the final reduced upper triangular matrix is still denoted by $\R$.

\begin{algorithm}[h!]
\caption{LLL reduction}   \label{a:LLL}
\begin{algorithmic}[1]
  \STATE compute the QR factorization \eqref{e:qr}
  \STATE set $\Z=\I_n$, $k=2$;
  \WHILE{$k\le n$}
   \STATE size reduce $r_{k-1,k}$ and update $\Z$
   \IF{$\delta\,  r_{k-1,k-1}^2>  r^2_{k-1,k}+r^2_{kk}$}
    \STATE permute and triangularize $\R$ and update $\Z$
    \STATE $k=k-1$, when $k>2$;
   \ELSE
    \FOR{$i=k-2,\dots,1$}
    \STATE size reduce $r_{ik}$ and update $\Z$
    \ENDFOR
    \STATE $k=k+1$;
   \ENDIF
  \ENDWHILE
\end{algorithmic}
\end{algorithm}

\section{Effects of lattice reduction on $P_{ZF}$ }\label{s:effects}

The following theorem gives a formula for computing the success probability of the ZF decoder $\x^{ZF}$, which is defined as:
\beq \label{e:ZF}
\x^{ZF}=\lfloor \R^{-1}\tby\rceil,
\eeq
for \eqref{e:ILSR}. The formula is equivalent to the one given in \cite{Teu98a}, which
considers a variant form of the ILS problem  \eqref{e:ILS}.

\begin{theorem} \label{t:prob}
Let $P_{ZF}$ denote the success probability  of the ZF decoder  $\x^{ZF}$ given in  \eqref{e:ZF}.
Then with $\bxi=\bmx \xi_1 & \xi_2 &\cdots &\xi_n\emx^T\in \mathbb{R}^{n}$, we have
\begin{align*}
P_{ZF}=&\Pr(\x^{ZF}=\hbx)= \frac{|\det \R|}{(2\pi\sigma^2)^{n/2}}\\
&\times\int_{-0.5}^{0.5}\cdots\int_{-0.5}^{0.5}\exp(-\frac{1}{2\sigma^2}\parallel\R\bxi\parallel^2_2)d\xi_1 \cdots d\xi_n.
\end{align*}
\end{theorem}

{\em Proof.} Let $\x^R=\R^{-1}\tby$, by \eqref{e:ZF},
$$
\x^{ZF}=\hbx\,\Leftrightarrow\,  -0.5\leq x_i^R-\hat{x}_i\leq0.5, \,i=1,2,\cdots, n.
$$
Since $\tby\sim \mathcal{N}(\R\hbx,\sigma^2 \I)$, we have
$$\x^R-\hbx\sim \mathcal{N}(0,\sigma^2(\R^T\R)^{-1}),$$
so the theorem holds.
 \ \ $\Box$

Since $P_{ZF}$ depends on $\R$, sometimes we write $P_{ZF}$ as $P_{ZF}(\R)$.
Effective methods of computing $P_{ZF}$ can be found in \cite{GasDS02} and references therein.

We can also define the corresponding ZF decoder $\z^{ZF}$ for \eqref{e:reduced}.
$\z^{ZF}$ can be used as an estimator of $\hbz \equiv \Z^{-1}\hbx$, or equivalently $\Z\z^{ZF}$ can be used to estimate $\hbx$.
In  \eqref{e:ILSR}, $\tby\sim {\cal N}(\R\hbx, \sigma^2\I)$.  It is easy to verify that
in \eqref{e:reduced}, $\bby \sim {\cal N}(\bbR\hbz, \sigma^2\I)$.

The following theorem shows that both SQRD and V-BLAST have no effect on $P_{ZF}$.

\begin{theorem} \label{t:perm}
Suppose that the ILS problem \eqref{e:ILSR} is transformed to the ILS problem \eqref{e:reduced}, where $\bbR$ is obtained by using SQRD or V-BLAST, then
\beqnn
\Pr(\x^{ZF}=\hbx)=\Pr(\z^{ZF}=\hbz).
\eeqnn
\end{theorem}
{\em Proof}. If SQRD or V-BLAST is used to do the reduction, then $\Z$ in \eqref{e:QRZ} is a permutation matrix. Therefore, we have
\begin{align}
 \label{e:xzzf}
\z^{ZF}&=\lfloor\bbR^{-1}\bby\rceil=\lfloor\Z^{-1}\R^{-1}\tby\rceil   \nonumber\\
&=\Z^{-1}\lfloor\R^{-1}\tby\rceil=\Z^{-1}\x^{ZF}.
\end{align}
Thus, $\z^{ZF}=\hat{\z}$ if and only if $\x^{ZF}=\hat{\x}$. Completes the proof.
\ \ $\Box$

We give some remarks here.
\begin{remark}
From Theorem \ref{t:perm} and its proof, one can easily see that the column permutations in the process of the LLL reduction {\em alone}
(i.e., the LLL-P defined in \cite{WenC14}) can not change $P_{ZF}$. Also, it is not hard to see that none of these column permutation strategies can
change the residual of the ZF decoder. In fact, by \eqref{e:QRZ} and \eqref{e:xzzf}, we have
\beqnn
\|\bby-\bbR\z^{ZF}\|_2=\|\bbQ^T\tby-\bbQ^T\R\Z\z^{ZF}\|_2=\|\tby-\R\x^{ZF}\|_2.
\eeqnn
\end{remark}

\begin{remark}
By the definition of the orthogonality defect, it is also easy to see that none of these column permutation strategies
can change the orthogonality defect.
\end{remark}

In the following we look at how size reductions affect $P_{ZF}$. Before the analysis we would like to give the following lemma.

\begin{lemma}\label{l:gauss}
Let $\zeta>0$, $\sigma\in \mathbb{R}$ and
$$
f(t)=\int_{-\zeta+t}^{\zeta+t}\exp(-\frac{x^2}{2\sigma^2})dx.
$$
Then $f(t)$ is a strict decreasing function when $t>0$.
\end{lemma}

{\em Proof.}
By some simple calculations, we have
\begin{align*}
f'(t)
&=\exp(-\frac{(\zeta+t)^2}{2\sigma^2})-\exp(-\frac{(\zeta-t)^2}{2\sigma^2})\\
&=\exp(-\frac{\zeta^2+t^2}{2\sigma^2})[\exp(-\frac{\zeta t}{\sigma^2})-\exp(\frac{\zeta t}{\sigma^2})].
\end{align*}
Clearly, $f'(t)<0$ when $t>0$. \ \ $\Box$

Using Lemma \ref{l:gauss}, we can obtain the following result.

\begin{lemma} \label{l:size}
Suppose that in \eqref{e:ILSR}, $\R=\bmx r_{11} & r_{12}\\ 0 & r_{22}\emx$ satisfies $|r_{12}|>\frac{1}{2}|r_{11}|$ and $\bbR$ in \eqref{e:reduced}  is
got by using only size reductions on $\R$, then
\beqnn
\Pr(\x^{ZF}=\hbx)<\Pr(\z^{ZF}=\hbz).
\eeqnn
\end{lemma}
{\em Proof}. Since $|\det(\R)|=|\det(\bbR)|$, by Theorem \ref{t:prob}, it is equivalent to show
\beq
\label{e:main}
P_1< P_2,
\eeq
where
\begin{align*}
&P_1=\int_{-0.5}^{0.5}\int_{-0.5}^{0.5}\exp(-\frac{1}{2\sigma^2}\parallel\R\bxi\parallel^2_2)d\xi_1d\xi_2, \nonumber\\
&P_2=\int_{-0.5}^{0.5}\int_{-0.5}^{0.5}\exp(-\frac{1}{2\sigma^2}\parallel\bbR\bxi\parallel^2_2)d\xi_1d\xi_2.
\end{align*}
Let $\L=\R^{-1}$ and $\zeta=\R\bxi$, then
\[
\L=\bmx \frac{1}{r_{11}} &-\frac{ r_{12}}{r_{11}r_{22}}\\ 0 & \frac{1}{r_{22}}\emx
\]
and $-\frac{1}{2}\leq\xi_1, \xi_2\leq \frac{1}{2}$ is equivalent to
\begin{align*}
-\frac{r_{11}}{2}+\frac{r_{12}}{r_{22}}\zeta_2\leq \zeta_1\leq\frac{r_{11}}{2}+\frac{r_{12}}{r_{22}}\zeta_2, \quad
-\frac{r_{22}}{2}\leq \zeta_2\leq\frac{r_{22}}{2}.
\end{align*}
Thus, by integration by substitution, we have
\begin{align*}
&\quad|\det(\R)|P_1\\
&=\int_{-\frac{r_{22}}{2}}^{\frac{r_{22}}{2}}\exp(-\frac{\zeta_2^2}{2\sigma^2})\int_{-\frac{r_{11}}{2}+\frac{r_{12}}{r_{22}}\zeta_2}
^{\frac{r_{11}}{2}+\frac{r_{12}}{r_{22}}\zeta_2}\exp(-\frac{\zeta_1^2}{2\sigma^2})d\zeta_1d\zeta_2\\
&=\int_{-\frac{r_{22}}{2}}^{\frac{r_{22}}{2}}\exp(-\frac{\zeta_2^2}{2\sigma^2})\int_{-\frac{r_{11}}{2}+|\frac{r_{12}}{r_{22}}\zeta_2|}
^{\frac{r_{11}}{2}+|\frac{r_{12}}{r_{22}}\zeta_2|}\exp(-\frac{\zeta_1^2}{2\sigma^2})d\zeta_1d\zeta_2.
\end{align*}
Since $\bbR$ is obtained by using only size reductions on $\R$,
$r_{11}=\bar{r}_{11}$ and $r_{22}=\bar{r}_{22}$. Thus, similarly, we have
\begin{align*}
&\quad|\det(\R)|P_2\\
=&\int_{-\frac{r_{22}}{2}}^{\frac{r_{22}}{2}}\exp(-\frac{\zeta_2^2}{2\sigma^2})\int_{-\frac{r_{11}}{2}+|\frac{\bar{r}_{12}}{r_{22}}\zeta_2|}
^{\frac{r_{11}}{2}+|\frac{\bar{r}_{12}}{r_{22}}\zeta_2|}\exp(-\frac{\zeta_1^2}{2\sigma^2})d\zeta_1d\zeta_2.
\end{align*}
Since $|r_{12}|>\frac{1}{2}|r_{11}|$, $\bar{r}_{12}< r_{12}$, by Lemma \ref{l:gauss}, for each fixed $\zeta_2$,
\begin{align*}
\int_{-\frac{r_{11}}{2}+|\frac{r_{12}}{r_{22}}\zeta_2|}
^{\frac{r_{11}}{2}+|\frac{r_{12}}{r_{22}}\zeta_2|}\exp(-\frac{\zeta_1^2}{2\sigma^2})d\zeta_1
<\int_{-\frac{r_{11}}{2}+|\frac{\bar{r}_{12}}{r_{22}}\zeta_2|}
^{\frac{r_{11}}{2}+|\frac{\bar{r}_{12}}{r_{22}}\zeta_2|}\exp(-\frac{\zeta_1^2}{2\sigma^2})d\zeta_1.
\end{align*}
Therefore, \eqref{e:main} holds.
\ \ $\Box$

From Theorem \ref{t:perm} and Lemma \ref{l:size}, one can easily obtain the following theorem.

\begin{theorem} \label{t:size}
Suppose that $n=2$ and the ILS problem \eqref{e:ILSR} is transformed to the ILS problem \eqref{e:reduced}, where $\bbR$ is obtained by using Algorithm \ref{a:LLL}, then
\beqnn
\Pr(\x^{ZF}=\hbx)\leq\Pr(\z^{ZF}=\hbz),
\eeqnn
where the equality holds if and only if there is no size reductions happens during the process of the LLL reduction.
Although column permutation {\em alone} can not change $P_{ZF}$, every size reduction can strictly improve $P_{ZF}$.
\end{theorem}

\begin{remark}
Although column permutation {\em alone} can not change $P_{ZF}$, if the column permutation can bring size reductions, then $P_{ZF}$ can be further improved.
We give the following example to illustrate it.
\bex
Let $\sigma=0.5$ and $\R=\bmx 4 & 9\\ 0 & 1\emx$.
We get $\bbR_1=\bmx 4 & 1\\ 0 & 1\emx$ if only size reductions are used to $\R$,
and we get $\bbR_2=\bmx \sqrt{2} & 0\\ 0 & 2\sqrt{2}\emx$ if the LLL reduction is used.
By Theorem \ref{t:prob}, we have $P_{ZF}(\R)=0.3413$, $P_{ZF}(\bbR_1)=0.6825$ and $P_{ZF}(\bbR_2)=0.8388$.
\eex
\end{remark}

Theorem \ref{t:size}  shows that the LLL reduction can always improve (not strictly) $P_{ZF}$ when $n=2$, so we can discuss how the parameter $\delta$ affects the improvement.
Suppose  that $\bbR_1$ and $ \bbR_2$ are obtained by applying Algorithm \ref{a:LLL} to $\A$
with $\delta=\delta_1$ and $\delta=\delta_2$, respectively and $\delta_1 < \delta_2$.
A natural question is what is the relation between $P_{ZF}(\bbR_1)$ and $P_{ZF}(\bbR_2)$?
The following theorem which is similar to \cite[Theorem 4]{ChaWX13} answers it.

\begin{theorem}\label{t:2delta}
Suppose  that $1/4<\delta_1 < \delta_2\leq 1$, and
$\bbR_1$ and $ \bbR_2$ are obtained by applying Algorithm \ref{a:LLL} to $\A\in \mathbb{R}^{m\times n}$
with $\delta=\delta_1$ and $\delta=\delta_2$, respectively.
If $n=2$,  then
\beq
P_{ZF}(\bbR_1)\leq P_{ZF}(\bbR_2). \label{eq:p12}
\eeq
 \end{theorem}

For the sake of reading, we follow the proof of \cite[Theorem 4]{ChaWX13} to prove it.

{\em Proof}.
Note that only two columns are involved in the reduction process and the value of $\delta$ only determines
when the process should terminate.
In the reduction process, the upper triangular matrix $\R$ either first becomes $\delta_1$-LLL reduced and then becomes $\delta_2$-LLL reduced
after some more size reductions or  becomes $\delta_1$-LLL reduced and $\delta_2$-LLL reduced at the same time.
Therefore, by Lemma \ref{l:size}, the strict inequality in \eqref{eq:p12} holds if further size reductions
are needed in the former case, otherwise, the equality in \eqref{eq:p12} holds.
\ \ $\Box$

\begin{remark}
Although the LLL reduction can improve $P_{ZF}$ when $n=2$, it may increase the residual of the ZF decoder.
The following example illustrates this.
\bex
Let $\R=\bmx 1 & 0.44\\ 0 & 0.28\emx$ and $\y=\bmx -0.7 \\ -0.24\emx$, after LLL reduction we have
$\bbR=\bmx 0.5215 & -0.1994\\ 0 & -0.5369\emx$ and $\bby=\bmx -0.7194 \\ 0.1733\emx$. After some simple
calculus, we have $\|\tby-\R\x^{ZF}\|_2=0.2631$ and $\|\bby-\bbR\z^{ZF}\|_2=0.3672$.
\eex
\end{remark}

\medskip

Although the LLL reduction can always improve (not strictly) $P_{ZF}$ when $n=2$, it may decrease $P_{ZF}$ if $n\geq 3$.
The following example confirms this.

\bex
\label{ex:counterE}
Let $\sigma=1$ and
$\R=\bmx 3 & 1.5 & 0\\ 0& 3 &-1.51\\0& 0& 3\emx,$
then after the LLL reduction, i.e., Algorithm \ref{a:LLL}, we get
$\bbR=\bmx 3 & 1.5 & 1.5\\ 0& 3 &1.49\\0& 0& 3\emx.$
Using Matlab, we can get $P_{ZF}(\R)=0.6105$ and $P_{ZF}(\bbR)=0.6030$.

With the aforementioned $\R$ and $n>3$, and define  $\A=\bmx \I_{n-3} & 0\\ 0& \R\emx$. Let $\bar{\A}$ be the
LLL reduced matrix of $\A$, then obviously $P_{ZF}(\A)>P_{ZF}(\bar{\A})$.
\eex

Although Example \ref{ex:counterE} shows that after the LLL reduction, $P_{ZF}$ may be decrease when $n\geq 3$,
generally speaking, $P_{ZF}$ will increase.
This is because $\bbR$ is reduced and $\bbR^{-1}\tby$ tends to have a "smaller" variance matrix.
Simulation results on this can be found in, e.g., \cite{GanM08}.

\section{Summary} \label{s:sum}

We have rigorously shown that none of SQRD, V-BLAST and LLL-P can change $P_{ZF}$.
We have also shown that the LLL reduction can always improve (not strictly)
$P_{ZF}$ when $n=2$, and analyzed how the parameter $\delta$ affect the enhancement of $P_{ZF}$ when $n=2$.
A counter example has also been given to show that the LLL reduction may decrease $P_{ZF}$ when $n\geq3$.

\section*{Acknowledgment}
We are  grateful to the anonymous referees for their valuable and thoughtful suggestions
which improve the quality of the paper a lot.

\bibliographystyle{IEEEtran}.
\bibliography{ref}

\end{document}